Original Paper

# Stochastic phenomena in a fiber Raman amplifier

*Vladimir Kalashnikov[1,5], Sergey V. Sergeyev[1,\*], Juan Diego Ania-Castanón[2],*
*Gunnar Jacobsen[3] and Sergei Popov[4]*

*Corresponding Author: s.sergeyev@aston.ac.uk

[1]Aston Institute of Photonic Technologies, Aston University, Aston Triangle, Birmingham,
B4 7ET, UK
[2] Instituto de Optica CSIC, Serrano 121, Madrid, 28006, Spain
[3]Acreo, Electrum 236, SE-16440, Kista, Sweden
[4]Royal Institute of Technology (KTH), SE-1640, Stockholm, Sweden
[5]Institute of Photonics, Vienna University of Technology, Gusshausstr. 27/387, Vienna, A-
1040, Austria

The interplay of such cornerstones of modern nonlinear fiber optics as a nonlinearity, stochasticity and polarization leads to variety of the noise induced instabilities including polarization attraction and escape phenomena harnessing of which is a key to unlocking the fiber optic systems specifications required in high resolution spectroscopy, metrology, biomedicine and telecommunications. Here, by using direct stochastic modeling, the mapping of interplay of the Raman scattering-based nonlinearity, the random birefringence of a fiber, and the pump-to-signal intensity noise transfer has been done in terms of the fiber Raman amplifier parameters, namely polarization mode dispersion, the relative intensity noise of the pump laser, fiber length, and the signal power. The obtained results reveal conditions for emergence of the random birefringence-induced resonance-like enhancement of the gain fluctuations (stochastic anti-resonance) accompanied by pulse broadening and rare events in the form of low power output signals having probability heavily deviated from the Gaussian distribution.

## 1. Introduction

Manakov equations as a versatile tool for modeling the interplay of the nonlinearity (Raman-, Brillouin-, Kerr-based etc.), random birefringence-based stochasticity and polarization in nonlinear fiber optics have recently addressed many challenges related to studying modulation and polarization instabilities, polarization properties of dissipative solitons, parametric and Raman amplifiers, etc. [1-15]. This type of equations can be obtained by averaging Maxwell equations over the fast randomly varying birefringence along the fiber length at the scale of the random birefringence correlation length (RBCL) [1-3]. Sergeyev and coworkers have suggested a modification of such averaging technique for studying the polarization properties of fiber Raman amplifiers (FRAs) [12-14], which has been recently justified by direct stochastic modeling [15]. The modified technique is based on accounting for the scale of the signal-to-pump states of polarization (SOP) interaction in addition to the scale of the RBCL. As a result, it reveals the presence of a stochastic anti-resonance (SAR) phenomenon, leading to resonance-like enhancement of the gain fluctuations for short fiber lengths of 5 km with increased polarization mode dispersion (PMD) parameter $D_p$ [12-18]. Modern ultra-long (>200 km) fiber Raman-based unrepeatered transmission schemes explore bi-directional pumping schemes [19] where forward pump has a major contribution to the PMD-dependent pump-to-signal relative intensity noise (RIN) transfer [5]. Thus, such pump configuration is very important in the context of studying the contribution of random birefringence-mediated stochastic properties of FRAs including SAR phenomenon into the RIN transfer [20, 21]. In this article for the first time, we use a direct stochastic modeling of pump and signal SOPs evolution along the fiber with accounting for pump depletion to get insight into the range of parameters (PMD, fiber length, signal power and RIN of the pump laser) where SAR affects FRAs performance. The separation of the deterministic and stochastic evolution gives us the opportunity of obtaining a library of stochastic trajectories. This allows for a substantial reduction in computational resource requirements (time and required memory) which is important in the context of telecommunication applications (in areas such as machine learning based modulation format recognition [22], linear and nonlinear transmission impairments mitigation [23] or stochastic digital backpropagation [24]) as well as fiber laser design (e.g. machine learning based self-adjustment of optimal laser parameters [25]).

We demonstrate that, though the fraction of the Raman gain caused by signal-pump SOPs interaction and the gain fluctuations caused by random birefringence decrease due to the longer lengths typical of FRA and the PMD parameter, SAR still can boost the pump-to-signal RIN transfer up to 10 dB for FRA lengths over 40 km and $D_p$ of 0.02 ps km$^{-1/2}$ whereas increased the pump RIN supresses the emergence of the rare events in the form of the low power signal pulses with the probability heavily deviated from the Gaussian distribution. By comparing the results of stochastic simulations with those obtained with two different analytical models it is possible to determine the margins of the analytical models' validity, which is of particular relevance for their use in the aforementioned applications [22-25], in which fast computing is paramount.

## 2. Stochastic vector models of fiber Raman amplifier

Our analysis is based on the direct numerical integration of the stochastic equations modeling vector Raman amplification taking into account the depletion of a pump co-propagating with a signal [1]. We neglect cross-phase and self-phase modulations (XPM and SPM) as well as group-delay and group velocity dispersion (GVD). The XPM and SPM can be omitted for pump powers $P_{in} < 1$ W, signal powers $s_0 < 10$ mW and $D_p > 0.01$ ps km$^{-1/2}$ [5, 6]. It has been estimated [5, 6] that GVD can be dropped out when the fiber length $L$ is much smaller than the dispersion length $L_d = T_p^2/|\beta_2|$. For the pulse duration $T_p = 2.5$ ps, $|\beta_2| = 5$ ps$^2$/km, we have $L_D > 100$ km. Thus, GVD can be neglected for $L < 50$ km [5, 6].

In the Jones representation [1, 2] and without taking into account group-delay dispersion and any nonlinear effect other than the Raman gain, the coupled Manakov-PMD equations are [3,4]:

$$\frac{d|A_s\rangle}{dz} = \left[\frac{g_R}{2}|A_p\rangle\langle A_p| - \alpha_s + b_s\left(\sigma_3\cos\theta + \sigma_1\sin\theta\right)\right]|A_s\rangle,$$

$$\frac{d|A_p\rangle}{dz} = \left[-\frac{g_R\omega_p}{2\omega_s}|A_s\rangle\langle A_s| - \alpha_p + b_p\left(\sigma_3\cos\theta + \sigma_1\sin\theta\right)\right]|A_p\rangle, \tag{1}$$

where $|A_{p,s}\rangle$ are the Jones vectors (time-dependent in a general case) for a signal ($s$-index) and a pump ($p$-index), respectively; $\sigma_1 = \begin{pmatrix} 0 & 1 \\ 1 & 0 \end{pmatrix}$, $\sigma_2 = \begin{pmatrix} 0 & -i \\ i & 0 \end{pmatrix}$ and $\sigma_3 = \begin{pmatrix} 1 & 0 \\ 0 & -1 \end{pmatrix}$ are the Pauli matrixes; the Raman gain as well as the loss coefficients for pump and signal are defined by $g_R$, $\alpha_p$ and $\alpha_s$, respectively, and $z$ is the propagation distance. The effect of fiber birefringence is defined by parameters $b_s = 2\pi/L_{bs}$ and $b_p = 2\pi/L_{bp}$ where $L_{bs}$ and $L_{bp}$ are the beat lengths for a signal and a pump at the frequencies $\omega_s$ and $\omega_p$, respectively. The stochastic birefringence is defined by the Wiener process $d\theta/dz = \zeta(z)$ so that $\langle\zeta(z)\rangle = 0$, $\langle\zeta(z)\zeta(z')\rangle = \sigma^2\delta(z-z')$, $\sigma^2 = 1/L_c$, $L_c$ is the coherence length for random birefringence, and $\theta$ is the orientation angle of birefringence vector in the equator plane of Poincaré sphere. The rotation

$$
\begin{aligned}
&\left|A_{p,s}\right\rangle = R\left|a_{p,s}\right\rangle, \\
&R = \begin{pmatrix} \cos(\theta/2) & \sin(\theta/2) \\ -\sin(\theta/2) & \cos(\theta/2) \end{pmatrix},
\end{aligned}
\tag{2}
$$

and the unitary transformation

$$
\begin{aligned}
&T = \begin{pmatrix} t_1 & t_2 \\ -t_2^* & t_1^* \end{pmatrix}, \frac{dT}{dz} = \Sigma T, \ \Sigma = \begin{pmatrix} ib_p & \zeta \\ -\zeta & -ib_p \end{pmatrix}, \\
&\left|U\right\rangle = T\left|a_s\right\rangle, \left|V\right\rangle = T\left|a_p\right\rangle,
\end{aligned}
\tag{3}
$$

result in the following equations for the new signal and pump Jones vectors:

$$
\begin{aligned}
&\frac{d\left|U\right\rangle}{dz} = \frac{g_R}{2}\left|V\right\rangle\left\langle V\middle|\middle|U\right\rangle - \alpha_s\left|U\right\rangle + i\left(b_s - b_p\right)\tilde{\sigma}_3\left|U\right\rangle, \\
&\frac{d\left|V\right\rangle}{dz} = \frac{g_R\omega_p}{2\omega_s}\left|U\right\rangle\left\langle U\middle|\middle|V\right\rangle - \alpha_p\left|V\right\rangle, \\
&\tilde{\sigma}_3 = T^+\sigma_3 T.
\end{aligned}
\tag{4}
$$

It is a common practice to use the Stokes representation instead of the Jones one for studying polarization properties of FRA [4-15]. The relation between the Stokes and Jones representations is [4-5]:

$$
\vec{S} = \left\langle U\middle|\vec{\sigma}\middle|U\right\rangle, \ \vec{P} = \left\langle V\middle|\vec{\sigma}\middle|V\right\rangle, \ \vec{\sigma} = \sigma_1\vec{i} + \sigma_2\vec{j} + \sigma_3\vec{k},
\tag{5}
$$

where $\vec{i}$, $\vec{j}$ and $\vec{k}$ are basis vectors [4, 5].

A remarkable property of the Jones representation in the form of Eqs. (4) is that such representation allows separating the stochastic part governed by the equation for the components of matrix $T$ in (3) from the field dynamics governed by (4). This separation simplifies calculations substantially, especially in the case when temporal effects like group-delay, its dispersion, and Kerr nonlinearity are taken into account.

The systems (3,4) have been solved by the 2.0-order stochastic scalar noise Runge-Kutta algorithm in Mathematica and MATLAB problem-solving environments (see Supporting Information). The stochastic equations (3) have to be understood in the Stratonovich sense. As was found this algorithm provides the best accuracy for reasonable step sizes $\Delta z = 10^{-3}\text{-}10^{-4}\min(L_c, L_b)$ (details can be found in Supporting Information). The results of our direct numerical simulations have been compared with those of the multi-scale averaging technique [12-15]. Such a technique allows obtaining the equations for averaged values:

$$\frac{d\langle s_0\rangle}{dz'}=\varepsilon_1\exp\left(-\varepsilon_2 z'\right)\langle x\rangle,\ \frac{d\langle x\rangle}{dz'}=\varepsilon_1\exp\left(-\varepsilon_2 z'\right)\langle s_0\rangle-\varepsilon_3\langle y\rangle,$$

$$\frac{d\langle s_0^2\rangle}{dz'}=2\varepsilon_1\exp\left(-\varepsilon_2 z'\right)\langle s_0 x\rangle,\ \frac{d\langle y\rangle}{dz'}=\varepsilon_3\left(\langle x\rangle-\langle\tilde{p}_1\tilde{s}_1\rangle\right)-\frac{L}{2L_c}\langle y\rangle,$$

$$\frac{d\langle y^2\rangle}{dz'}=2\varepsilon_3\left(\langle xy\rangle-\langle y\rangle\langle\tilde{p}_1\tilde{s}_1\rangle\right)-\frac{L}{L_c}\left(\langle y^2\rangle-\langle u^2\rangle\right),$$

$$\frac{d\langle x^2\rangle}{dz'}=2\varepsilon_1\exp\left(-\varepsilon_2 z'\right)\langle s_0 x\rangle-2\varepsilon_3\langle xy\rangle,$$

(6)

$$\frac{d\langle xy\rangle}{dz'}=\varepsilon_1\exp\left(-\varepsilon_2 z'\right)\langle s_0 y\rangle+\varepsilon_3\left(\langle x^2\rangle-\langle x\rangle\langle\tilde{p}_1\tilde{s}_1\rangle\right)-\frac{L}{2L_c}\langle xy\rangle,$$

$$\frac{d\langle s_0 y\rangle}{dz'}=\varepsilon_1\exp\left(-\varepsilon_2 z'\right)\langle xy\rangle-\frac{L}{2L_c}\langle s_0 y\rangle+$$

$$+\varepsilon_3\left(\langle s_0 x\rangle-\langle y^2\rangle-\langle s_0\rangle\langle\tilde{p}_1\tilde{s}_1\rangle\right),$$

$$\frac{d\langle s_0 x\rangle}{dz'}=\varepsilon_1\exp\left(-\varepsilon_2 z'\right)\left(\langle s_0^2\rangle+\langle x_0^2\rangle\right)-\varepsilon_3\langle s_0 y\rangle,$$

$$\frac{d\langle u^2\rangle}{dz'}=\frac{L}{L_c}\left(\langle y^2\rangle-\langle u^2\rangle\right),$$

where $L$ is the fiber length, $z'=z/L$, $\langle x\rangle=\langle\vec{s}\cdot\vec{p}\rangle$ is the average projection of a signal SOP to a pump SOP; $\langle y\rangle=\langle\tilde{p}_3\tilde{s}_2-\tilde{p}_2\tilde{s}_3\rangle$, $\langle\tilde{p}_1\tilde{s}_1\rangle=\tilde{p}_1(0)\tilde{s}_1(0)\exp\left(-z'L/L_c\right)$, $\langle u\rangle=\langle\tilde{p}_3\tilde{s}_1-\tilde{p}_1\tilde{s}_3\rangle$, $\varepsilon_1=g_R P(0)L/2$, $\varepsilon_2=\alpha_s L$ and $\varepsilon_3=2\pi L\left(\omega_p/\omega_s-1\right)/L_{bp}$. Here normalization $\vec{S}=s_o\vec{s}\exp\left(\int\limits_0^L\left(g_R p_0(z')/2-\alpha_s z'\right)dz'\right)$ is used. $P_{in}$ is an input pump power, $s_0=\left|\vec{S}\right|$, $p_0=\left|\vec{P}\right|$, $\vec{s}$ and $\vec{p}$ are the unit vectors defining the signal and pump SOPs, respectively; $\tilde{\vec{s}}$ and $\tilde{\vec{p}}$ are the unit SOP vectors in the reference frame where the birefringence vector is defined as $\vec{W}_p=\left(2b_p,0,0\right)$ [8].

The limit of $L\gg L_c$ is understandable from the results of the averaging of Eqs. (3,4) over the stochastic birefringence (below such a limit will be called as the "Manakov limit", see Supporting Information). The result in Stokes representation can be expressed in the following form:

$$\frac{d\langle\vec{S}\rangle}{dz}=\frac{g_R}{2}\left(\langle|\vec{P}|\rangle\langle\vec{S}\rangle+\langle|\vec{S}|\rangle\langle\vec{P}\rangle\right)-\alpha_s\langle\vec{S}\rangle+$$

$$+\left(b_s-b_p\right)\begin{pmatrix}0\\-\langle S_3\rangle\\\langle S_2\rangle\end{pmatrix}e^{-2\sigma^2 z},$$

(7)

$$\frac{d\langle\vec{P}\rangle}{dz}=-\frac{g_R\omega_p}{2\omega_s}\left(\langle|\vec{P}|\rangle\langle\vec{S}\rangle+\langle|\vec{S}|\rangle\langle\vec{P}\rangle\right)-\alpha_p\langle\vec{P}\rangle,$$

The parameters of the average maximum gain $\langle G \rangle$, its relative standard deviation $\sigma_G$ and the polarization dependent gain ($PDG$) are defined as follows [14, 15]:

$$\langle G \rangle = 10 \log \left( \frac{\left\langle \left| \vec{S}(L) \right| \right\rangle}{\left| \vec{S}(0) \right|} \right), \sigma_G = \sqrt{\frac{\left\langle \left| \vec{S}(L) \right|^2 \right\rangle}{\left\langle \left| \vec{S}(L) \right| \right\rangle^2} - 1},$$

$$PDG = 10 \log \left( \frac{s_{0,\max}(L)}{s_{0,\min}(L)} \right). \tag{8}$$

Here $s_{0,\max}(L)$ and $s_{0,\min}(L)$ correspond to the signal powers at $z = L$ for the initially collinear and orthogonal signal-power SOPs, respectively. These parameters will be matter at issue in this work.

Unlike the averaged equations derived in [7], Eqs. (7) have been derived with the help of the unitary transformation shown in Eqs. (3) which preserves the length of the pump and signal SOP vectors as well as the scalar and vector products. As a result, the evolution of the signal SOP includes a term accounting for the relative rotation of the signal SOP relatively the pump SOP. This approach differs from that of Kozlov and co-workers [7], which applied an unitary transformation to the pump and signal SOPs to exclude both the pump and signal SOP fluctuations due to random birefringence, leading to a set of equations different from (7). The stimulated Raman scattering and the cross-phase modulation introduce a coupling between the pump and signal SOPs and, therefore, the vector and scalar products are not preserved in the equations obtained in [7].

## 3. Results and discussion

Maximum average gain $\langle G \rangle$ and its relative standard deviation $\sigma_G$ are shown in Figs. 1 and 2 as functions of the fiber length $L$ and the PMD parameter $D_p \equiv 2\lambda_s \sqrt{L_c} / c L_{bs}$. Since the Raman gain depends on the relative pump-signal SOPs, the maximum gain corresponds to initially collinear pump and signals SOPs, i.e. $\vec{s} = \vec{p} = (1,0,0)$. The results presented have been obtained from Eqs. (3,4) with averaging over 100 independent stochastic trajectories. One can see that the average gain increases with the PMD decrease ($D_p \lesssim 0.01$ ps km$^{-1/2}$) as well as with the moderate growth of fiber length, achieving maximum value for $L \gtrsim 10$ km (Fig. 1). Such behavior can be explained by the polarization pulling (trapping) when the pump SOP pulls the signal SOP so that a Raman gain medium acts as an effective polarizer (i.e., $\langle x \rangle \to 1$, see Eqs. (6)) [12,14-19]. Since a long interaction distance enhances a pulling, the gain grows with $L$. However for extra-long fibers ($L \gtrsim 16$ km in our case), pump depletion in combination with fiber loss cause a gradual decrease of $\langle G \rangle$. Simultaneously, the relative depolarization of pump and signal SOPs prevents pulling for large PMD and this effect is most pronounced within some interval of fiber lengths (in the vicinity of $L \approx 6$ km in our case). Both limits of small and large $L$ suppress the sensitivity of $\langle G \rangle$ to PMD. As follows from Fig.1, such suppression corresponds to the limit of $L \gg L_c$ that is understandable from Eqs. (7), which demonstrate the reduction of birefringence effects with distance and PMD parameter decrease (i.e., decrease of $L_c$).

The influence of stochastic birefringence on the relative deviation of gain is illustrated in Fig. 2 which shows resonant enhancement of gain fluctuations within a confined region of fiber lengths and PMDs. One can see a maximum of gain relative standard deviation in the vicinity of $L \approx 6$ km and $D_p \approx 0.02$ ps km$^{-1/2}$. We connect this phenomenon with the manifestation of stochastic anti-resonance (SAR) when oscillations of relative pump-signal SOPs induced by fiber birefringence enhance the effect of stochastic birefringence if $L_{bp}$, $L_{bs}$ approaches $L_c$. Such an enhancement of stochastization distinguishes this phenomenon from the stochastic resonance occurring when the noise increases the correlation between input and output signals and the signal-to-noise ratio passes through a maximum [20,21].

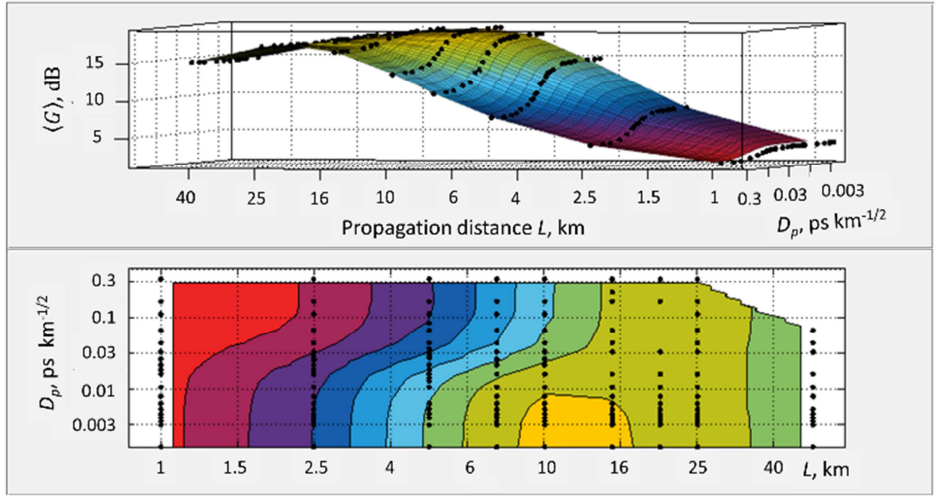

**Figure 1** Dependence of average maximum gain coefficient $\langle G \rangle$ on a fiber length $L$ and a PMD parameter $D_p$ (points correspond to numerical data fitted by 3D-surface). Correlation length $L_c$ is of 100 m, Raman gain coefficient $g_R$ is of 0.8 W$^{-1}$ km$^{-1}$, $P_{in} = 1$ W, $\left| \vec{S}(0) \right| = 10$ mW, and loss coefficients $\alpha_p \approx \alpha_s$ for pump and signal are of 0.2 dB km$^{-1}$. Signal and pump SOPs are collinear initially: $\vec{s} = \vec{p} = (1, 0, 0)$.

The SAR appearing for some value of PMD was interpreted in [15] as an activated escape from the polarization trapping quantified in terms of a steep dropping of the corresponding Kramer length and Hurst parameter in the vicinity of the maximum of the gain fluctuations. On the other hand, the resonant enhancement of standard deviation for some interval of $L$ can be easily interpreted from Figure 3. For short propagation lengths, the stochastic trajectories of signal power diverge insignificantly because the trajectory wandering has no sufficient time to develop. However, the divergence of trajectories grows with distance. At large distances ($\gtrsim 16$ km in the case shown), gain depletion begins to contribute to the gain evolution. As a result, a bundle of stochastic trajectories squeezes because the low-power trajectories continue to grow but high-power trajectories sink due to pump depletion and fiber loss. Thus, the negative feedback owing to pump depletion inhibits the noise caused by stochastic fiber birefringence (Figs. 2, 3). Simultaneously, a trace of "unbundled" low-power trajectories remains (Fig. 3) that defines a peculiarity of signal statistics (see below).

The passive negative feedback induced by pump depletion affects the temporal profiles of both signal and pump (inset in Fig. 3, see Supporting Information as well), namely, it forms a hole on the pump profile (continuous wave, initially) and transforms a Gaussian signal pulse into super-Gaussian one with the subsequent growth of the pulse width $\tau$. Dependence of the averaged signal width $<\tau>$ on the PMD parameter is shown in Fig. 4. One can see a noticeable growth of $<\tau>$ caused by polarization pulling in the $D_p \to 0$ limit (black solid curve in Fig. 4). Simultaneously, a nonlinearity induced by pump depletion transfers the power fluctuations into the pulse width ones (red dashed curve in Fig. 4) so that SAR affects the pulse width, as well (compare the narrow probability distributions in the Manakov and "diffusion" limits with that on the SAR vertex, insets in Fig. 4).

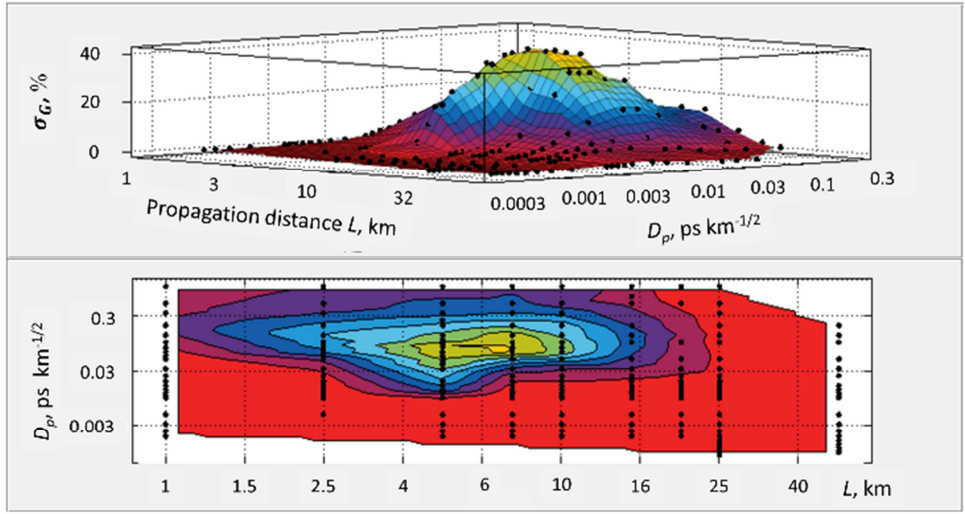

**Figure 2** Relative standard deviation of the maximum gain coefficient corresponding to Fig. 1.

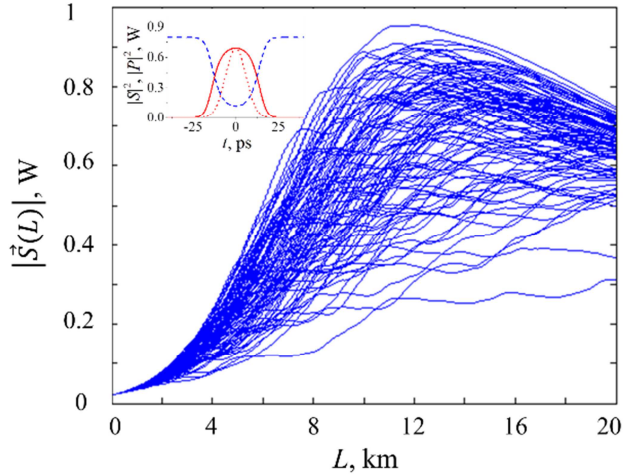

**Figure 3** Signal power dependence on propagation distance (1000 stochastic trajectories are shown). The inset shows the temporal profiles of a signal (solid red curve) and a pump (blue dashed curve) for one trajectory at $L$=20 km; the initial signal pulse is Gaussian with FHWM of 20 ps (dotted red curve scaled by a factor corresponding to $G$=18.4 dB); the initial pump is continuous-wave. Parameters: $D_p$ = 0.026 ps km$^{-1/2}$, other parameters correspond to Fig. 1.

One has to note that the condition of $L > > L_c$ is met for SAR so that the result of Eqs. (7) is valid only for stronger inequality. Equations (7) are valid in the limits of $L/L_c \to \infty$ for an arbitrary $D_p$ or $D_p \to 0$ for an arbitrary $L$ or for both limits that mean the validity of a statement $L/L_c \to \infty$ and/or $D_p \to 0$ for the Manakov limit described by Eqs. (7) (Fig. 5).

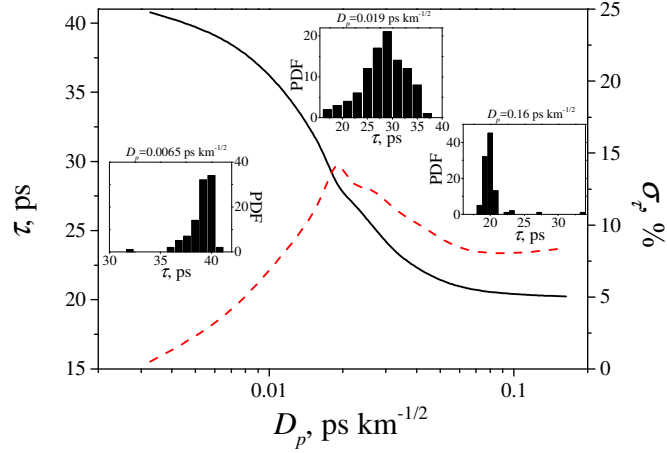

**Figure 4**. Dependences of the average signal pulse width $\langle \tau \rangle$ (black solid curve) and its relative deviation $\sigma_\tau$ (red dashed curve) on the PMD parameter. Insets show the pulse width probability distributions for $D_p = 0.0065$, 0.019, and 0.16 ps km$^{-1/2}$. Input signal pulse is Gaussian with the FWHM of 20 ps and the peak power of 10 mW; $L$=20 km; signal and pump SOPs are collinear initially.

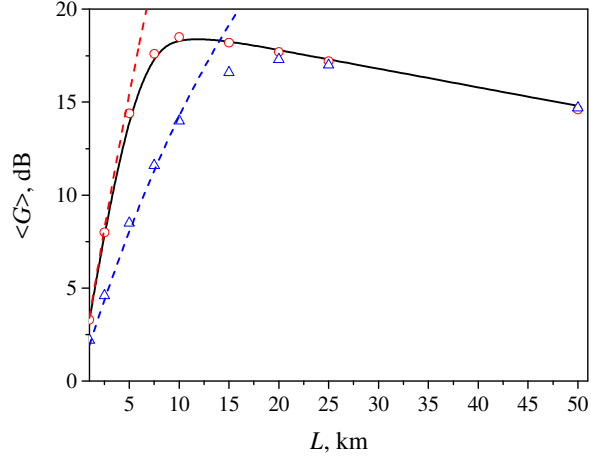

**Figure 5** Dependence of the maximum $\langle G \rangle$ on fiber length $L$ in the Manakov limit of Eqs. (8) (solid curve), and for the exact model of Eqs. (1): $D_p$ = 0.001 ps km$^{-1/2}$ (circles) and 0.065 ps km$^{-1/2}$ (triangles). Red and blue dashed curves correspond to the results obtained from Eqs. (7) for $D_p$ = 0.001 ps km$^{-1/2}$ and 0.065 ps km$^{-1/2}$, respectively. Signal and pump SOPs are collinear initially: $\vec{s} = \vec{p} = (1,0,0)$.

With PMD growth, a deviation of an exact $\langle G \rangle$ (triangles, Fig. 5) from the Manakov limit (solid curve) appears for fiber lengths near SAR ( $L \approx 5 \div 10$ km). One has to note that the multi-scale averaging technique, leading to Eqs. (6) (dashed curves in Fig. 5), provides an excellent approximation to exact numerical results for both small and large PMDs at distances when depletion does not contribute substantially. An important advance of this technique is that it describes statistical properties such as gain relative standard deviation, relative signal gain, etc. [15]

When $L/L_c \to \infty$ and $D_p \to 0$, the relative standard deviation of the maximum gain coefficient vanishes that corresponds to the domination of polarization pulling. The relative standard deviation of gain is maximum in the region of SAR (Fig. 2), and the limit of large PMD and small fiber lengths is characterized by decreasing but sufficiently large relative standard deviation of gain that results from decorrelation of rapid relative rotations of pump-signal SOPs (i.e., their average relative projection $\langle x \rangle \to 0$). <G> is minimal here and is defined by a "diffusion" limit (Fig. 5 for small $L$) [14].

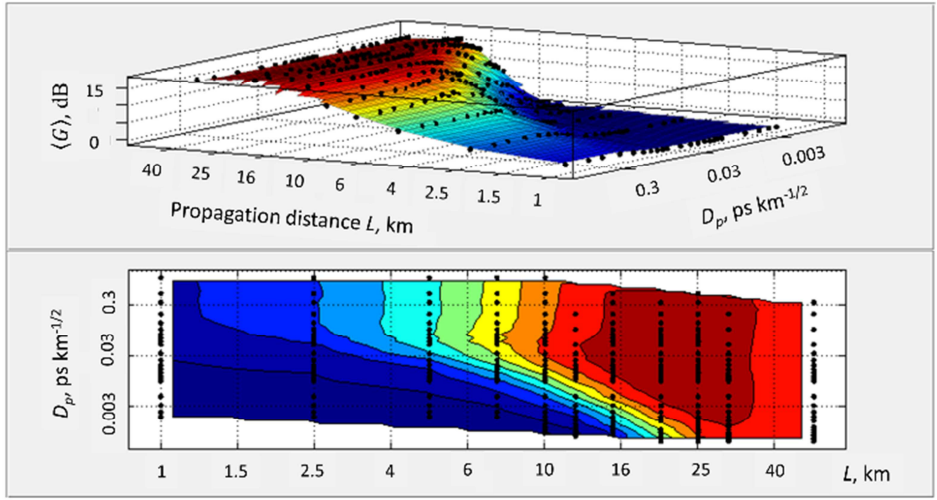

**Figure 6** Dependence of average minimal gain coefficient <G> on a fiber length $L$ and a PMD parameter $D_p$. Parameters correspond to Figure 1. Signal and pump SOPs are orthogonal initially: $\vec{s} = (-1, 0, 0)$ and $\vec{p} = (1, 0, 0)$.

The case of minimal gain corresponding to initially orthogonal pump and signal SOPs (i.e. $\vec{s} = (-1, 0, 0)$ and $\vec{p} = (1, 0, 0)$) is illustrated in Figs. 6 and 7. One can see a threshold-like growth of average gain with the growth of fiber length and PMD parameter. Such growth can be interpreted in the terms of escape from a metastable state [16-18] of $\langle x \rangle = -1$ due to fluctuations of SOPs induced by stochastic birefringence. Figure 7 demonstrates an extremal enhancement of such fluctuations in the vicinity of threshold-growth of the average gain coefficient. Since the interaction length of signal-pump SOPs increases with a fiber length, the polarization pulling increases, as well. As a result, the average gain coefficient grows, and its fluctuations are suppressed. However, despite the case of Fig. 2, the relative standard deviation becomes maximal with minimization of PMD for a sufficiently large $L$ (Fig. 7). Such a shift of SAR into the region of smaller PMDs and larger fiber lengths can be explained by the deceleration of SOP evolution to a polarization pulling state for the case of initially orthogonal signal-pump SOPs with the decrease of $D_p$.

Figure 8 displays the *PDG* parameter. One can see the *PDG* suppression with increasing fiber length and PMD parameter. The first tendency results from the enhancement of polarization pulling with the growth of signal-pump SOPs interaction length. The second tendency results from the intensive decorrelation of signal-pump SOPs with the decrease of beat length. Simultaneously, *PDG* becomes extremely enhanced for small PMD in the vicinity of $L \approx 10$ km, which correlates with the conjunction of SAR positions for minimal and maximal gains (see Figs. 2 and 7).

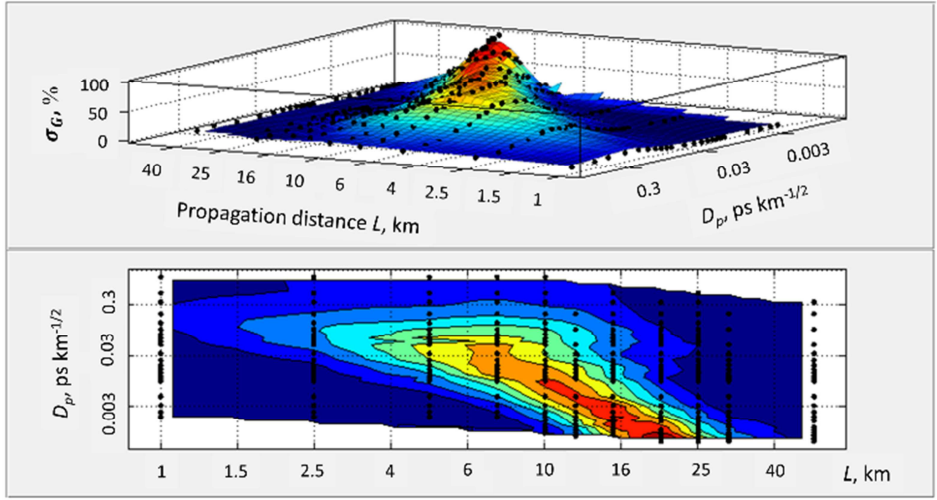

**Figure 7** Relative standard deviation of the minimum gain coefficient corresponding to Figure 6.

As modern fiber Raman-based telecommunication systems support high-level signal powers (up to 100 mW) [19], it is of interest to consider the influence of input signal power on maximum gain and its relative standard deviation. Though this value is beyond of the margins of the Eqs. (3, 4) validity, the model based on Eqs. (3, 4) can be used for qualitative evaluation of the signal power influence on the averaged gain and gain fluctuations. Figure 9 demonstrates the dependencies of <G> and $\sigma_G$ on the PMD parameter for different levels of $\left|\vec{S}(0)\right|$. One can see a monotonic reduction of gain with input signal power due to pump depletion. Such a reduction increases for small PMDs due to the effect of polarization pulling, which causes more effective Raman amplification and, as a consequence, more intensive pump depletion. Simultaneously, the relative standard deviation of gain decreases with the $\left|\vec{S}(0)\right|$-growth, but the position of SAR does not change appreciably. One may assume that the reduction of relative standard deviation with the signal power growth results from a stronger polarization pulling which synchronizes the evolution of pump-signal SOPs.

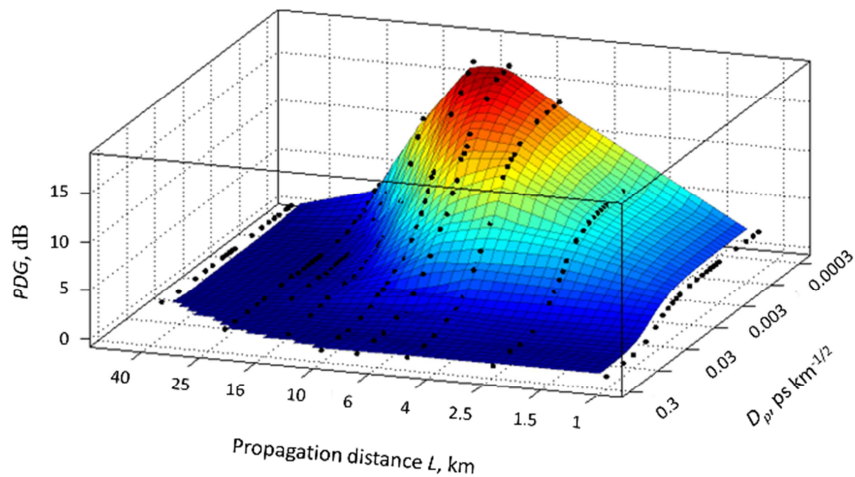

**Figure 8**. *PDG* for the parameters corresponding to Figures 1, 6.

As was pointed out above, the gain noise caused by stochastic birefringence depends resonantly on both PMD and propagation length. In real-world amplifiers, there are additional noise sources due to the pump power and SOP fluctuations. Here, we consider the effect of pump power fluctuations on a signal in relation to the PMD parameter and the fiber length. Figure 10 demonstrates the mean gain standard deviations with (dashed curves) and without (solid curves) additive Gaussian pump noise. One can see that stochastic birefringence dominates for large PMD, i.e. around the region of SAR. Here, the integral pump-noise transfer function defined as

$$H_{\text{int}} = 10 \log \left( \frac{\int_0^\infty RIN_s(f) df}{\int_0^\infty RIN_p(f) df} \right) = 10 \log \left( \sigma_G^2 / \sigma_P^2 \right), \tag{9}$$

(here $\sigma_P$ is the relative standard deviation for noisy pump and $RIN_{s,p}$ are the corresponding relative intensity noises at frequency $f$ [25]) takes the maximum value (dots connected by dotted curves in Fig. 10). The decrease of PMD reduces both $\sigma_G$ and $H_{\text{int}}$, however, they do not vanish for the noisy pump but tend to some constant value for $D_p < <1$ and $L/L_c > >1$. It is clear that the contribution of stochastic birefringence is suppressed by polarization pulling in the $D_p \to 0$ limit so that only pump noise contribution remains. Simultaneously, pump noise is suppressed by $L$-growth due to pump depletion (see Fig. 3). However, such suppression becomes saturated with $L$-growth (Fig. 10) that is testified by non-Gaussian statistics (Figs 11 and 12c, d).

The statistical properties of signal power without input pump noise are illustrated in Fig. 11. In the diffusion limit, i.e. in the limit of large PMD, Raman amplification is almost polarization independent due to fast relative rotation of pump and signal SOPs. As a result, the signal PDF close to Gaussian and is narrow, but with a visible substructure due to residual polarization effects (Figure 11, a). Approaching to SAR broadens PDF and enhances its visible substructure so that distribution becomes rather "nonparametric" [26] (Fig. 11b). Transition to small PMD or/and large propagation distances (Manakov limit) results in a non-Gaussian statistic which is close to "extreme value" one [27] with the prolonged low power tail (Figures 11c, d). Figure 3 illustrates the corresponding contribution of low-power stochastic trajectories. It is obvious that the cutting edge in this distribution is defined by polarization pulling providing maximum gain. Nevertheless, some sustained fluctuations of the relative SOPs around this state remain.

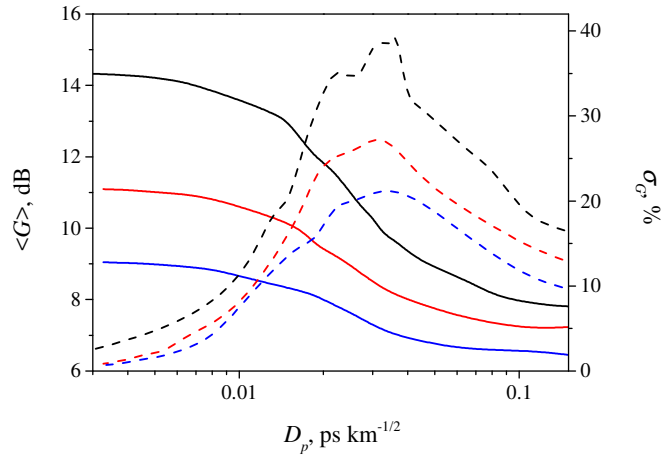

**Figure 9** Maximum gain <G> (solid curves) and its relative standard deviation $\sigma_G$ (dashed curves) for different input signal powers: 20 mW (black), 50 mW (red) and 100 mW (blue). $L$=5 km, other parameters correspond to Fig. 1.

Pump noise modifies the signal probability distributions, especially in the region of small PMD (Fig. 12). One can see from Figs. 12 (a,b), that the modifications are not substantial for the large PMD and the relatively small propagation distances where the SAR contribution prevails. However, a Gaussian input noise modifies distributions for small PMD or/and large $L$ (see Figs. 12 (c,d)) so that they become rather a Gaussian (or Burr-like in the case of (d)) than extreme one.

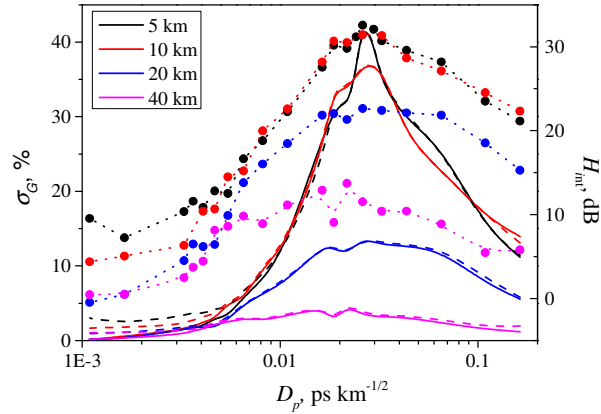

**Figure 10.** Dependencies of the relative standard deviation $\sigma_G$ (solid and dashed curves concern the noiseless and noisy pumps, respectively) and the integral pump-noise transfer function $H_{\text{int}}$ (dots connected by dotted curves) on the PMD parameter for different fiber lengths: 5 km (black), 10 km (red), 20 km (blue), and 40 km (magenta). Pump power fluctuations are described by a Gaussian noise with the standard deviation of 1%. Other parameters correspond to Figure 1.

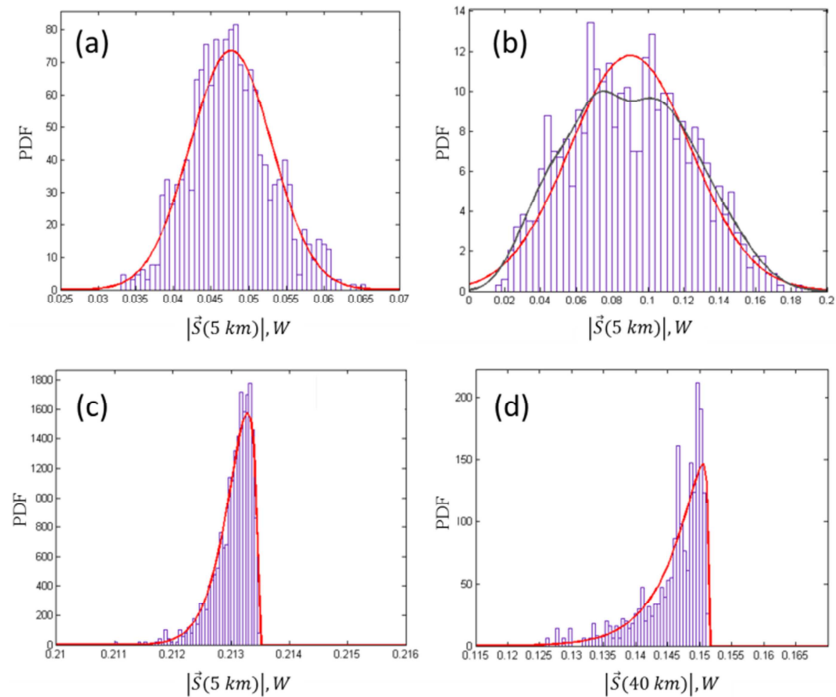

**Figure 11**. Probability distributions of signal power for 1000 samples without pump noise. $D_p = 0.16$ (a), 0.026 (b,d), and 0.001 ps km$^{-1/2}$ (c) and $L$=5 (a-c) and 40 km (d). Fitting curves correspond to Gaussian distribution (red curves in (a), (b)), nonparametric distribution (blue curve in (b)), and generalized extreme distribution (red curves in (c), (d)) obtained with the help of MATLAB statistics toolbox. Other parameters correspond to Figure 1.

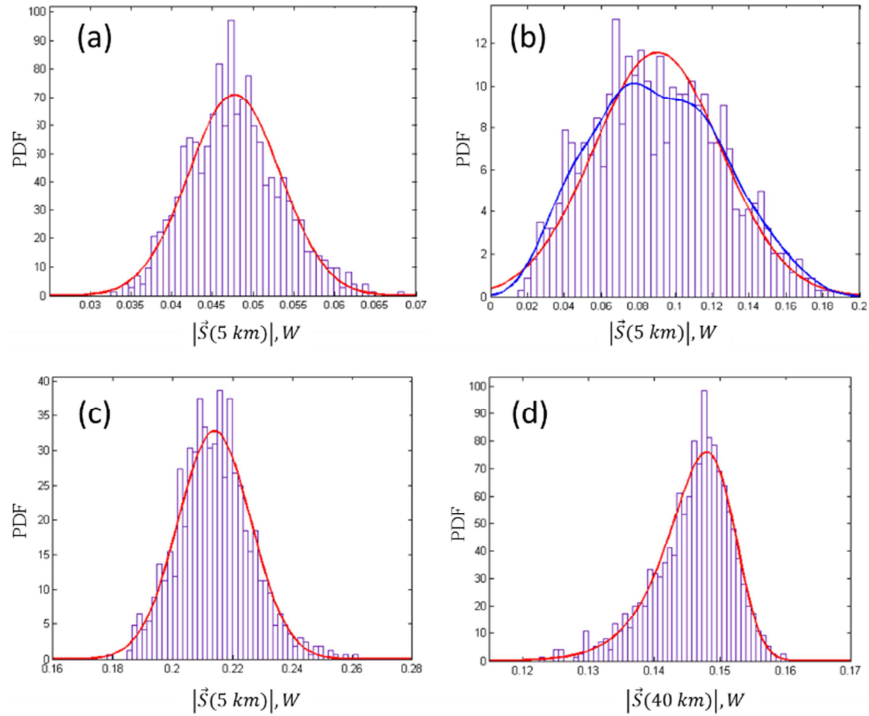

**Figure 12** Probability distributions of signal power for 1000 samples with Gaussian pump noise. $D_p$ = 0.16 (a), 0.026 (b,d), and 0.001 ps/km$^{-1/2}$ (c) and $L$=5 (a-c) and 40 km (d). Fitting curves correspond to Gaussian distribution (red curves in (a), (b) and (c)), nonparametric distribution (blue curve in (b)), and Burr distribution (red curve in (d)) obtained with the help of MATLAB statistics toolbox. Other parameters correspond to Fig. 1.

## 4. Conclusion

Using direct stochastic simulation of coupled Manakov-PMD equations, we evaluate the contribution of the stochastic anti-resonance phenomenon on fiber Raman amplifier gain, gain fluctuations and pump-to-signal intensity noise transfer as a function of PMD parameter, fiber length and the signal input power. We have found that although stochastic anti-resonance impact on the averaged gain and the random birefringence-mediated gain fluctuations becomes negligible with increased length and PMD parameter, it still contributes substantially to the pump-to-signal RIN transfer up to 10 dB for FRA lengths over 40 km and $D_p$ of 0.02 ps km$^{-1/2}$. In addition, the SAR leads to the pulse broadening and emergence of the rare events taking the form of low output power pulses with the probability heavily deviated from the Gaussian distribution. With the help of the separation of the deterministic and stochastic evolution we have obtained a library of stochastic trajectories that allowed us to substantially decrease computational time and required memory. This improvement in computational efficiency can find potential applications in telecommunications for machine learning based modulation format recognition [22], linear and nonlinear transmission impairments mitigation [23], stochastic digital backpropagation [24] as well as in fiber lasers engineering for machine learning based optimal laser parameters self-adjustment [25]. Comparison of the results of stochastic simulations with the results for two analytical models has helped to determine the margins of the analytical models as follows: (i) the model based on Eqs. (6) is using two-scale averaging and neglects the pump depletion and so is valid for all PMD values, the small input signal powers (<1 mW) and short fiber lengths of less than 6 km, (ii) the model based on Eqs. (7) is using single-scale averaging is valid in the limit of the fiber lengths of $L>25$ km or small PMD values. The margins can help to justify the application of the analytical models in the aforementioned limit cases to reduce substantially the computational time required for machine learning based problems solving [22-25]. Including the fluctuations of the pump SOP, pump-signal walk-off, group-delay dispersion and Kerr-nonlinearity will introduce some corrections to signal-noise transfer for large fiber lengths but the analysis of these effects is beyond the scope of the present article. Nevertheless, one has to note that Eqs. (5) admit a straightforward generalization taking into account these effects that will be an object of our forthcoming analysis.

**Acknowledgements**    This work has been funded through grants FP7-PEOPLE-2012-IAPP (project GRIFFON, No. 324391), Leverhulme Trust (Grant ref: RPG-2014-304), Spanish MINECO TEC2015-71127-C2-1-R (project ANOMALOS), and Austrian Science Fund (FWF project P24916-N27).

Received: ((will be filled in by the editorial staff))
Revised: ((will be filled in by the editorial staff))
Published online: ((will be filled in by the editorial staff))

**Keywords**: Stochastic processes, Nonlinear fiber optics, Polarization Phenomena, Fiber optic amplifiers, Raman effect.

Supporting Information

## Stochastic phenomena in a fiber Raman amplifier

*Vladimir Kalashnikov[1,5], Sergey V. Sergeyev[1,*], Juan Diego Ania-Castanón[2], Gunnar Jacobsen[3] and Sergei Popov[4]*

*Corresponding Author: s.sergeyev@aston.ac.uk

[1]Aston Institute of Photonic Technologies, Aston University, Aston Triangle, Birmingham, B4 7ET, UK
[2] Instituto de Optica CSIC, Serrano 121, Madrid, 28006, Spain
[3]Acreo, Electrum 236, SE-16440, Kista, Sweden
[4]Royal Institute of Technology (KTH), SE-1640, Stockholm, Sweden
[5]Institute of Photonics, Vienna University of Technology, Gusshausstr. 27/387, Vienna, A-1040, Austria

**This document provides supporting information to "Stochastic phenomena in a fiber Raman amplifier." We describe numerical and averaging techniques underlying our results.**

## 1. Numerical technique

The structure of Eqs. (4,5) allows splitting the numerical simulations into two independent parts notably i) independent solution of a system of stochastic ordinary differential equations (SDE) (4) that gives a "library" of stochastic material parameters for ii) simulation of field evolution (in Jones representation) based on Eqs. (5) with independent stochastic coefficients. One has to note that Eqs. (5) can be generalized with taking into account group-delay, group-delay dispersion, and Kerr nonlinearity so that the resulting model becomes a split system of SDE + "nonlinear partial differential equations with independent stochastic coefficients." As the solution of SDE is most challenging task, a careful choice of numerical technique is required. The algorithm convergence can be controlled by following property of the unitary $T$-matrix:

$$\left| t_1 \left( z \right) \right|^2 + \left| t_2 \left( z \right) \right|^2 = 1, \, t_1 \left( 0 \right) = 1, \, t_2 \left( 0 \right) = 0, \tag{ES1}$$

and unitarity has to persist during evolution by virtue of unitarity of $\Sigma$. Figure S1 shows a behavior of invariant (ES1) calculated with the help of *Wolfram Mathematica* 10.0 using *StratanovichProcess* and *RandomFunction* procedures, and four different methods: "*Milstein*" and "*Stochastic Runge-Kutta*" (Figure S1, (1)), as well as "*Stochastic Runge-Kutta Scalar Noise*" and "*Kloeden-Platen-Schurz*" (Figure S1, (2)). One can see that last two algorithms preserve (ES1), and they were used in our simulations.

To accelerate calculations substantially, we used *MATLAB*. In this case, the fast and good convergence was provided by the *Weak Order 2 Stochastic Runge-Kutta Method* [1]:

$$Y_0 = X_0,$$
$$U = Y_n + A\big(Y_n\big)\delta + B\big(Y_n\big)\Delta W_n,$$
$$U_\pm = Y_n + A\big(Y_n\big)\delta \pm B\big(Y_n\big)\sqrt{\delta},$$
$$Y_{n+1} = Y_n + \frac{\delta}{2}\Big[A\big(U\big) + A\big(Y_n\big)\Big] + \frac{\Delta W_n}{4}\Big[B\big(U_+\big) + B\big(U_-\big) + 2B\big(Y_n\big)\Big] +$$
$$+ \frac{\Big(\big(\Delta W_n\big)^2 - \delta\Big)}{4\sqrt{\delta}}\Big[B\big(U_+\big) - B\big(U_-\big)\Big], \tag{ES2}$$

where the Stratonovich type SDE is expressed formally in the form of

$$X\big(z\big) = A\big(z, X\big)dz + B\big(z, X\big) \circ dW\big(z\big). \tag{ES3}$$

In Eqs. (ES2), $\delta = z_{n+1} - z_n$, $\Delta W = W\big(z_{n+1}\big) - W\big(z_n\big)$, $W(z)$ is a Wiener process, and $X$, $A$, $B$ can be multi-component.

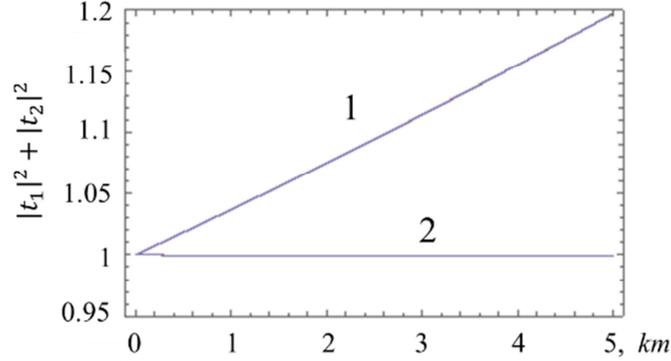

**Figure S1.** Evolution of $\big|t_1\big(z\big)\big|^2 + \big|t_2\big(z\big)\big|^2$-invariant from Eq. (4) simulated by the Milstein and Stochastic Runge-Kutta (1) as well as the Stochastic Runge-Kutta Scalar Noise and Kloeden-Platen-Schutz (2) algorithms.

## 2. Averaging technique and Manakov limit

Let us express $\tilde{\sigma}_3$ in Eq. (5) in the following form:

$$\tilde{\sigma}_3 = \begin{pmatrix} X_1 & X_4^* \\ X_4 & -X_1 \end{pmatrix}, \tag{ES4}$$

where a new six-component vector $X$ is introduced:

$$X_1 = \left| t_1 \right|^2 - \left| t_2 \right|^2, \; X_2 = -\left( t_1 t_2 + t_1^* t_2^* \right), \; X_3 = i \left( t_1 t_2 - t_1^* t_2^* \right),$$

$$X_4 = 2 i t_1 t_2^*, \; X_5 = t_1^2 - t_2^{*2}, \; X_6 = -i \left( t_1^2 + t_2^{*2} \right),$$

$$(ES5)$$

which obeys to SDE (ES3) with

$$A(X) = \left( 0, -2 b_p X_3, 2 b_p X_2, 0, -2 b_p X_6, 2 b_p X_5 \right)^T,$$

$$B(X) = \left( 2 X_2, -2 X_1, 0, 2 X_5, -2 X_4, 0 \right)^T.$$

$$(ES6)$$

Splitting of the initial system (2) into the stochastic material and field parts simplifies an averaging over the stochastic birefringence [2]. Applying the Dunkin formula and the Stratonovich generator $\hat{\Gamma}$ [3-5]:

$$\frac{d \left\langle f(\Omega_l) \right\rangle}{dz} = \hat{\Gamma} \left[ f(X_k) \right],$$

$$\hat{\Gamma} = \sum_{k=1}^{6} A_k \frac{\partial}{\partial X_k} + \frac{1}{2 L_c} \sum_{k=1}^{6} \sum_{l=1}^{6} \left( B_k B_l + B_l \frac{\partial B_k}{\partial X_l} \frac{\partial}{\partial X_k} \right)$$

$$(ES7)$$

($f$ is some function of $X$) results in the following groups of equations for averaged $\left\langle f(X) \right\rangle$:

$$\frac{d \left\langle X_1 \right\rangle}{dz} = -2 \sigma^2 \left\langle X_1 \right\rangle, \quad \frac{d \left\langle X_2 \right\rangle}{dz} = -2 b_p \left\langle X_3 \right\rangle - 2 \sigma^2 \left\langle X_2 \right\rangle,$$

$$\frac{d \left\langle X_3 \right\rangle}{dz} = 2 b_p \left\langle X_2 \right\rangle, \quad \frac{d \left\langle X_4 \right\rangle}{dz} = -2 \sigma^2 \left\langle X_4 \right\rangle,$$

$$\frac{d \left\langle X_5 \right\rangle}{dz} = -2 b_p \left\langle X_6 \right\rangle - 2 \sigma^2 \left\langle X_5 \right\rangle, \quad \frac{d \left\langle X_6 \right\rangle}{dz} = 2 b_p \left\langle X_5 \right\rangle,$$

$$\left( \left\langle X_1 \right\rangle, \left\langle X_2 \right\rangle, \left\langle X_3 \right\rangle \left\langle X_4 \right\rangle, \left\langle X_5 \right\rangle, \left\langle X_6 \right\rangle \right) \Big|_{z=0} = (1,0,0,0,0,0).$$

$$(ES8)$$

$$\frac{d \left\langle X_1^2 \right\rangle}{dz} = 4 \sigma^2 \left( \left\langle X_2^2 \right\rangle - \left\langle X_1^2 \right\rangle \right),$$

$$\frac{d \left\langle X_2^2 \right\rangle}{dz} = 4 \sigma^2 \left( \left\langle X_1^2 \right\rangle - \left\langle X_2^2 \right\rangle \right) - 4 b_p \left\langle X_2 X_3 \right\rangle,$$

$$\frac{d \left\langle X_3^2 \right\rangle}{dz} = 4 b_p \left\langle X_2 X_3 \right\rangle,$$

$$\frac{d \left\langle X_2 X_3 \right\rangle}{dz} = 2 b_p \left( \left\langle X_2^2 \right\rangle - \left\langle X_3^2 \right\rangle \right) - 2 \sigma^2 \left\langle X_2 X_3 \right\rangle,$$

$$\left( \left\langle X_1^2 \right\rangle, \left\langle X_2^2 \right\rangle, \left\langle X_3^2 \right\rangle, \left\langle X_2 X_3 \right\rangle \right) \Big|_{z=0} = (1,0,0,0).$$

$$(ES9)$$

$$\frac{d\langle X_4^2\rangle}{dz}=4\sigma^2\left(\langle X_5^2\rangle-\langle X_4^2\rangle\right),$$

$$\frac{d\langle X_5^2\rangle}{dz}=4\sigma^2\left(\langle X_4^2\rangle-\langle X_5^2\rangle\right)-4b_p\langle X_5 X_6\rangle,$$

$$\frac{d\langle X_6^2\rangle}{dz}=4b_p\langle X_5 X_6\rangle,$$

$$\frac{d\langle X_5 X_6\rangle}{dz}=2b_p\left(\langle X_5^2\rangle-\langle X_6^2\rangle\right)-2\sigma^2\langle X_5 X_6\rangle,$$

$$\left(\langle X_4^2\rangle,\langle X_5^2\rangle,\langle X_6^2\rangle,\langle X_5 X_6\rangle\right)\Big|_{z=0}=(0,1,-1,-i).$$

(ES10)

$$\frac{d\langle |X_4|^2\rangle}{dz}=4\sigma^2\left(\langle |X_5|^2\rangle-\langle |X_4|^2\rangle\right),$$

$$\frac{d\langle |X_5|^2\rangle}{dz}=4\sigma^2\left(\langle |X_4|^2\rangle-\langle |X_5|^2\rangle\right)-2b_p\left(\langle X_5 X_6^*\rangle+\langle X_6 X_5^*\rangle\right),$$

$$\frac{d}{dz}\left(\langle X_5 X_6^*\rangle+\langle X_6 X_5^*\rangle\right)=$$
$$=4b_p\left(\langle |X_5|^2\rangle-\langle |X_6|^2\rangle\right)-2\sigma^2\left(\langle X_5 X_6^*\rangle+\langle X_6 X_5^*\rangle\right),$$

$$\left(\langle |X_4|^2\rangle,\langle |X_5|^2\rangle,\langle X_5 X_6^*\rangle,\langle X_5 X_6^*\rangle+\langle X_6 X_5^*\rangle\right)\Big|_{z=0}=(0,1,1,0).$$

(ES11)

$$\frac{d\langle X_3 X_6\rangle}{dz}=2b_p\left(\langle X_2 X_6\rangle+\langle X_3 X_5\rangle\right),$$

$$\frac{d\langle X_2 X_5\rangle}{dz}=-2b_p\left(\langle X_2 X_6\rangle+\langle X_3 X_5\rangle\right)-4\sigma^2\left(\langle X_2 X_5\rangle-\langle X_1 X_4\rangle\right),$$

$$\frac{d\langle X_1 X_4\rangle}{dz}=4\sigma^2\left(\langle X_2 X_5\rangle-\langle X_1 X_4\rangle\right),$$

$$\frac{d}{dz}\left(\langle X_2 X_6\rangle+\langle X_3 X_5\rangle\right)=$$
$$=4b_p\left(\langle X_2 X_5\rangle-\langle X_3 X_6\rangle\right)-2\sigma^2\left(\langle X_2 X_6\rangle+\langle X_3 X_5\rangle\right),$$

$$\left(\langle X_3 X_6\rangle,\langle X_2 X_6\rangle+\langle X_3 X_5\rangle,\langle X_2 X_5\rangle,\langle X_1 X_4\rangle\right)\Big|_{z=0}=(0,0,0,0).$$

(ES12)

$$\frac{d\left\langle X_3 X_6^* \right\rangle}{dz} = 2b_p \left( \left\langle X_2 X_6^* \right\rangle + \left\langle X_3 X_5^* \right\rangle \right),$$

$$\frac{d\left\langle X_2 X_6^* \right\rangle}{dz} = 2b_p \left( \left\langle X_2 X_5^* \right\rangle - \left\langle X_3 X_6^* \right\rangle \right) - 2\sigma^2 \left\langle X_2 X_6^* \right\rangle,$$

$$\frac{d\left\langle X_2 X_5^* \right\rangle}{dz} = -2b_p \left( \left\langle X_2 X_6^* \right\rangle + \left\langle X_3 X_5^* \right\rangle \right) - 4\sigma^2 \left( \left\langle X_2 X_5^* \right\rangle - \left\langle X_1 X_4^* \right\rangle \right),$$

$$\frac{d\left\langle X_1 X_4^* \right\rangle}{dz} = 4\sigma^2 \left( \left\langle X_2 X_5^* \right\rangle - \left\langle X_1 X_4^* \right\rangle \right),$$

$$\frac{d\left\langle X_3 X_5^* \right\rangle}{dz} = 2b_p \left( \left\langle X_2 X_5^* \right\rangle - \left\langle X_3 X_6^* \right\rangle \right) - 2\sigma^2 \left\langle X_3 X_5^* \right\rangle,$$

$$\left( \left\langle X_3 X_6^* \right\rangle, \left\langle X_2 X_6^* \right\rangle, \left\langle X_2 X_5^* \right\rangle, \left\langle X_1 X_4^* \right\rangle, \left\langle X_3 X_5^* \right\rangle \right) \Big|_{z=0} = \left( 0,0,0,0,0 \right). \tag{ES13}$$

A group of Eqs. (ES8) has a trivial solution $\left\langle X_1 \right\rangle = \exp\left( -2\sigma^2 z \right)$ which leads to the "Manakov limit" equations (8) in the Stokes representation.

### 3. Pump depletion and signal/pulse temporal profiles

The real-world signal in Eqs. (4) is a pulse (i.e., $t$-dependent) whereas pump is initially continuous-wave (CW, i.e. $t$-independent). Eqs. (4) allow the direct extension of dimensionality including the $t$-coordinate [6]:

$$\frac{d|U\rangle}{dz} = \frac{g_R}{2} |V\rangle \langle V||U\rangle - \alpha_s |U\rangle + i\left( b_s - b_p \right) \tilde{\sigma}_3 |U\rangle + i\beta_s \tilde{\sigma}_3 \frac{\partial |U\rangle}{\partial t},$$

$$\frac{d|V\rangle}{dz} = \frac{g_R \omega_p}{2\omega_s} |U\rangle \langle U||V\rangle - \alpha_p |V\rangle + i\beta_p \tilde{\sigma}_3 \frac{\partial |V\rangle}{\partial t}, \tag{ES14}$$

where group-delay are defined by $\beta_{s,p} = \lambda_{s,p} b_{s,p} / 2\pi c$.

It is clear that pump depletion will modify both pump and signal profiles during evolution so that a passive (nonlinear) negative feedback appears. Such a feedback causes the signal pulse broadening, flattens pulse shape, and creates a gap in the pump CW-profile (Figure S2).

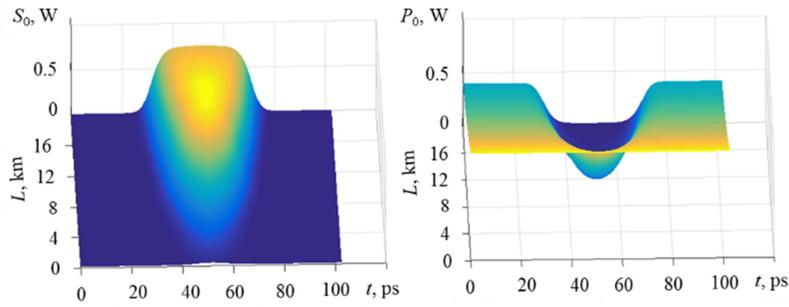

Figure S2. Signal (left) and pump (right) profiles along one stochastic trajectory for $D_p$=0.01 ps$^2$ km$^{-1/2}$. Initial signal pulse is Gaussian with 20 ps FWHM and 10 mW peak power; the initial pump is CW with 1 W power. The group-delays are not included.

Figure S2 presents an evolution along one stochastic trajectory of (ES14) defined by $\tilde{\sigma}_3$. The signals at $L$=20 km for six different trajectories are shown in Figure S3 for the same parameters. One can see that the pulse profile can become nontrivial due to the coupling of different polarization components.

Calculations suggest that impact of group-delay for $L$=20 km and the initial pulse width of even 5 ps is negligible.

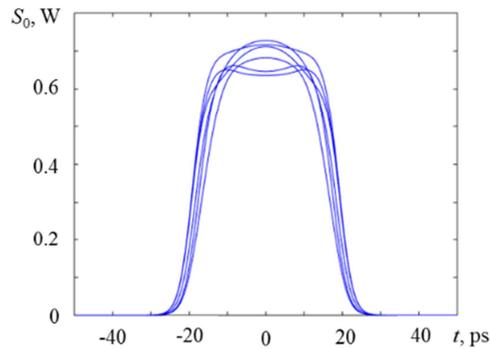

Figure S3. Final ($L$=20 km) signal profiles corresponding to six stochastic trajectories for $D_p$=0.01 ps$^2$ km$^{-1/2}$. Initial signal pulse is Gaussian with 20 ps FWHM and 10 mW peak power; the initial pump is CW with 1 W power. The group-delays are not included.

The relevant Mathematica and MATLAB codes, as well as the library of pre-calculated stochastic trajectories for (ES14) can be found in [7].